\documentclass{emulateapj}
\usepackage{graphicx}

\newcommand{\ltsima} {$\; \buildrel < \over \sim \;$}
\newcommand{\gtsima} {$\; \buildrel > \over \sim \;$}
\newcommand{\lta} {\lower.5ex\hbox{\ltsima}}
\newcommand{\gta} {\lower.5ex\hbox{\gtsima}}

\newcommand{\lsim}{\raisebox{-.4ex}{$\stackrel{<}{\scriptstyle \sim}$}}

\begin{document}
%\maketitle
%
%
\title {Disk-outflow coupling: Energetics around spinning 
black holes}

\author
{Debbijoy Bhattacharya\altaffilmark{1}, Shubhrangshu Ghosh\altaffilmark{2}, Banibrata Mukhopadhyay\altaffilmark{3}}
\affil{Astronomy and Astrophysics Programme, Department of Physics, Indian Institute of Science, Bangalore-560012, India.}

%\altaffiltext{1}{Astronomy and Astrophysics Program, Department of Physics,
%Indian Institute of Science, Bangalore 560012, India; bm@physics.iisc.ernet.in} 
\altaffiltext{1}{E-mail: debbijoy@physics.iisc.ernet.in}
\altaffiltext{2}{E-mail: sghosh@physics.iisc.ernet.in}
\altaffiltext{3}{E-mail: bm@physics.iisc.ernet.in}

%\author{Bhattacharya et al.}
%   {Debbijoy Bhattacharya$^{1}$ \thanks{E-mail: debbijoy@physics.iisc.ernet.in}, Shubhrangshu Ghosh$^{1}$ \thanks{E-mail: sghosh@physics.iisc.ernet.in}, 
%   Banibrata Mukhopadhyay$^{1}$\thanks{E-mail: bm@physics.iisc.ernet.in} \\
%  $^1$ Astronomy and Astrophysics Programme, Department of Physics, Indian Institute of Science, Bangalore-560012, India.  \\
%  }

%\pagerange{\pageref{firstpage}--\pageref{lastpage}} \pubyear{2009}

%\maketitle

%\label{firstpage}

\begin{abstract}

The mechanism by which outflows and plausible jets are driven from black
hole systems, still remains observationally elusive. Notwithstanding, several observational
evidences and deeper theoretical 
insights reveal that accretion and outflow/jet are strongly correlated.  
We model an 
advective disk-outflow coupled dynamics, incorporating
explicitly the vertical flux. Inter-connecting
dynamics of outflow and accretion essentially upholds the
conservation laws. We investigate the properties of the disk-outflow surface and 
its strong dependence on the rotation parameter of the black hole. 
The energetics of the disk-outflow strongly depend on the mass, accretion rate and spin of the black holes. 
The model clearly shows that the outflow power extracted from the disk increases strongly with the 
spin of the black hole, 
inferring that the power of the observed astrophysical jets has a proportional 
correspondence with the spin of the central object.
% The measured bolometric luminosity of on-axis jets as seen in  BL Lacs or Flat Spectrum Radio Quasars (FSRQs), which mostly emanates from the part of the jet compared to the part of the disk.
In case of blazars (BL Lacs and Flat Spectrum Radio Quasars), most of their emission are believed to be originated from their jets. It is observed that BL Lacs are relatively low luminous than Flat Spectrum Radio Quasars (FSRQs). The luminosity might be linked to the power of the jet, which in turn reflects that the nuclear regions of the BL Lac objects have a relatively low spinning black hole compared to that in the case of FSRQ. If the extreme gravity is the source 
to power strong outflows and jets, then the spin of the black hole, perhaps, might be the fundamental parameter to account for the observed astrophysical processes in an accretion powered system.

%\date{}
\end{abstract}

\maketitle 

%%%%%%%%%%%%%%%%%%%%%%%%%%%%%%%%%%%%%%%%%%%%%%%%%%%%%%%%%%%%%%%%

\section{Introduction}

High resolution observations show strong outflows and jets in black hole accreting systems, both in active galactic nuclei (AGNs) or quasars (Begelman et al. 1984; Mirabel 2003) and 
microquasars (SS433, GRS~1915+105) [Margon 1984; Mirabel \& Rodriguez 1994, 1998]. Extragalactic radio sources show evidence of strong jets in 
the vicinity of spinning black holes (Meier et al. 2001; Meier 2002). Outflows are also observed in neutron star 
low mass X-ray binaries (LMXBs) (Fender et al. 2004; Migliari \& Fender, 2006) and also in young stellar objects (Mundt 1985). 
It has been argued (Ghosh \& Mukhopadhyay 2009; Ghosh et al. 2010; hereinafter GM09, G10 respectively, and references therein) 
that outflows and jets are more prone 
to emanate from strong advective accretion flows; the said paradigm is more susceptible for super-Eddington and sub-Eddington 
accretion flows. However, the exact (global if any) mechanism of formation of the jet, its collimation, acceleration, composition 
from the accretion powered systems still remain inconclusive. 

Extensive 
works have been pursued on the origin of outflow/jet, since the 
pioneering work of Blandford \&
Payne (1982) (e.g. Pudritz \& Norman 1986; Contopoulos 1995; Ostriker 1997), where the authors used a self-similar approach to demonstrate that the 
poloidal component of the magnetic field can be seemingly used to launch outflowing 
matter from the disk. The formation of the jet is directly related to the efficacy of extraction of 
angular momentum and energy from the accretion disk. Physical understanding of the hydromagnetic outflows from 
disks has been developed from magnetohydrodynamic (MHD) simulations (Shibata \& Uchida 1986; Ustyogova et al. 1999; 
De Villiers et al. 2005; Hawley \& Krolik 2006) both in non-relativistic as well as in relativistic 
regimes, mostly in the Keplerian paradigm. However, the large Lorentz factors as well as Faraday rotation measures suggest that the 
observed VLBI jets in quasars and active galaxies are in the Poynting flux regime (Homan et al. 2001; Lyutikov 2006). 
On the other front, strong radiation pressure can serve as a different 
mechanism to effuse outflows/jets. This is likely to occur when the accretion rate is super-Eddington 
or super-critical (Fabrika 2004; Begelman et al. 2006; GM09,G10 and references therein) and the accretion disk is precisely ``radiation trapped'' as in 
ultra-luminous X-ray (ULX) sources. It has been confirmed by radiation hydrodynamic simulation at super-critical
accretion rate (Okuda et al. 2009) as well to explain luminosity and mass outflow rate of relativistic
outflows from SS433. The under-luminous 
accreting sources, having high sub-critical accretion flow, were explained by an advection dominated accretion flow (ADAF) model (Narayan \& Yi 1994). The promising 
outcome of this model lies in the large positive value of the Bernoulli parameter because of the small radiative energy
loss. It leads to conceive that the gas in the inflowing disk is susceptible to escape, leading to strong unbounded flows in the form of 
outflows and jets. This also signifies that even in the absence of magnetic field and radiation pressure, outflows are plausible from strongly advective accretion 
flow if the system is allowed to perturb. Nevertheless, the definitive understanding of the origin of outflows/jets is sill unknown. 
It remains one of
the most compelling problems in high energy astrophysics.  

Whatever might be the reason for the origin of outflow and then jet from the disk, one aspect is however definite that strong outflows producing relativistic jets are powered 
by extreme gravity. Although seems paradoxical, the strength and the length scale of observed astrophysical jets vary directly with the strength of the 
central gravitating potential. Observationally, it is evident that strong outflows 
and relativistic
jets are more powerful in observed AGNs and quasars, harboring supermassive black 
holes, compared to that in black hole X-ray binaries (XRBs). 
In addition, the jets observed in AGNs and quasars  
have greater length scale compared to that seen in stellar mass black hole systems. 
%Therefore, in any theoretical 
%modeling of the outflow and jet, relativistic effect of the central potential should be taken 
%into consideration. 
One of the most important signatures of relativistic 
gravitation is the spin of the central object. It is presumably believed that the spin (practically specific angular momentum) of the neutron star is less than that of a black hole, 
and thus the observed jet from black holes is much stronger and powerful than that of neutron star sources. In early, Blandford \& Znajek (1977) demonstrated that if there is a magnetic field associated with 
the black hole due to threading of magnetic field lines from the disk and the angular momentum of the Kerr black hole is large enough, then the energy and the angular 
momentum can be extracted from the underlying black hole by a purely electromagnetic mechanism, which can thus be expected to power the jet in an AGN. These imply that the 
spin might play a significant role in powering jets, both in microquasars and in AGNs, rather, 
it can act as a fundamental parameter in an accretion powered system. 

Most of the studies of accretion disk and studies of related outflow/jet have been evolved separately, assuming these two apparently to be dissimilar objects. However, several 
observational inferences (for details see GM09, G10 and references therein) and improved understanding of accretion flow and outflow reveal  
that accretion and outflow/jet are strongly correlated. The unifying scheme of disk and outflow is essentially governed by conservation laws; conservations 
of matter, energy and momentum. Hence, in modeling the accretion and outflow simultaneously in any accretion powered system, following aspects should be taken into 
account: (1) the effect of relativistic central gravitational potential including its spin, (2) proper mechanism of origin of outflow/jet from the disk, (3) appropriate 
hydrodynamic equations (considering that the accreting gas be treated as a continuum fluid), capturing the information about the 
intrinsic coupling between inflow and outflow which are governed by the conservation laws in a strongly advective paradigm. 

Recently, GM09, G10 made 
an endeavor to explore the dynamics of the accretion-induced outflow 
around black holes/compact objects in a 2.5-dimensional paradigm. The authors 
formulated the disk-outflow coupled model in a more self-consistent way by 
solving a complete set of coupled partial differential hydrodynamic equations in 
a general advective regime through a self-similar approach in an axisymmetric, cylindrical coordinate system. They explicitly incorporated 
the information of the vertical flux in their model. However, they restricted their study to Newtonian regime, thus negating the requisite effect of general-relativity, 
especially the effect of spin of the central object. Based on the model, the authors computed the mass outflow rate and the power extracted by the outflow from 
the disk self-consistently, without proposing any prior relation between the inflow and outflow. 

In the present paper, we propose a new model for the accretion-induced outflow by extending the work of GM09, G10, by incorporating the general relativistic effect of the central potential without limiting ourselves to a self-similar regime. As the definitive mechanism of launching of outflows/jets 
is still evasive, we do not embrace any specific mechanism of outflow like inclusion of the magnetic field into our model equations. Nevertheless, the importance of the 
magnetic field can not be, in principle, discarded to explain the launching and collimation of jets off the accretion disk. Here, owing to our inability to solve coupled partial 
differential MHD equations in a 2.5-dimensional advective regime, we neglect the influence of the magnetic field in our model equations. However, the implicit coupling between the inflow and outflow, dictated by the conservation equations, have 
been taken into account appropriately. We arrange our paper as follows. In the next section, we present the formulation of our model. The section 3 describes the computational 
procedure to solve the model equations of the accretion-induced outflow. In sections 4 and 5, we study the dynamics and the energetics of the flow respectively. Finally, we 
end up in section 6 with a summary and discussion.

\section{Modeling the correlated disk-outflow system}

We formulate the disk-outflow coupled model by considering a 
geometrically thick accretion 
disk, which is strongly advective as strong outflows/jets are more likely to eject from a thick/puffed up 
region of the accretion flow (GM09,G10). The vertical flow is explicitly included 
in the system. The basic features of the model are similar to that in GM09. 
We adopt the cylindrical coordinate system to describe a steady, axisymmetric 
accretion flow. The dynamical flow parameters, namely, radial velocity ($v_r$), specific angular momentum ($\lambda$), vertical velocity or outflow velocity ($ v_z$), adiabatic sound speed ($c_s$), mass density ($\rho$) and pressure ($P$) depend both on radial and vertical coordinates. We have already highlighted the importance of the spin of black hole to power the outflow/jet and its presumed relative effect on the observation of various AGN classes. Thus we have included the effect of the spin in our model. As the spin of the black hole is a signature of pure general relativity, i.e., Einstein's gravitation, its effect on the accretion flow, especially in the inner region of the disk, is mimicked approximately with the use of pseudo-general-relativistic or pseudo-Newtonian potential (PNP). Because of the disk-outflow system to be geometrically thick and the flow to be 2.5-dimensional (not confined to the equatorial plane), we use the PNP of Ghosh \& Mukhopadhyay (2007), which is a pseudo-Newtonian vector potential, 
to capture the inner disk properties of the accretion flow around a Kerr black hole approximately.

At the first instant we neglect the effect of viscosity in our system. One of the reasons behind it is the unavailability of the effective computational technique to solve coupled partial differential viscous hydrodynamic conservation equations for the compressible flow. To make the inviscid assumption more arguable, it can also be noted that the angular momentum transport in the accretion flow, for which the necessity of turbulent viscosity is invoked, can alternatively take place purely by outflow. At this extreme end, the outflow extracts angular momentum from the disk allowing the matter to get accreted towards the black hole and hence the inviscid  assumption can be adopted.  This is in essence similar to the Blandford \& Payne (1982) mechanism to
extract energy and angular momentum from the magnetized disk, where the extraction of the angular momentum 
and energy is essentially done by the outflow, and not due to the viscous dissipation.  
The outflow originates from just above the equatorial plane of the disk and this lower boundary is maintained at $z=0$, when $v_z=0$, unlike other work where the outflow is hypothesized to effuse out from the disk surface (e.g. Xie \& Yuan 2008). Our model is effectively valid only in the region where disk and outflow are coupled, i.e. the region from where essentially the outflow is emanated from the accretion flow. Hence our study will remain confined within this predefined region. Further, we neglect the contribution of the magnetic field as 
argued in \S 1.

We circumvent the idea of vertical integration of the flow equations. The validity and the reliability of the height integrated equations is normally gratifying in the geometrically thin limit. In that circumstances, the flow velocities are likely to be more or less independent of the disk scale-height, which is not the case of the present paradigm of interest. We further consider that the disk to be non-self-gravitating, assuming that the mass of the disk to be much less than that of the black hole. The radial and vertical coordinates are expressed in units of $GM/c^2$, flow velocities in $c$, time in $GM/c^3$ and the specific angular momentum in the unit of $GM/c$. Here $G$, $M$ and $c$ are gravitational constant, mass of the black hole and speed of light respectively. The steady state, axisymmetric disk-outflow coupled equations in cylindrical geometry in the inviscid limit are then as follows.

(a) Mass transfer:
%\begin{eqnarray}
%\frac{\partial \rho}{\partial t} + 
%\frac{\partial (\rho v_j )}{\partial x_j}= 0 .
%\label{1a}
%\end{eqnarray}
\begin{eqnarray}
\frac{1}{r} \frac{\partial}{\partial r} (r \rho v_r) \,+ \, \frac{\partial}{\partial z} (\rho v_z) =  0, 
\label{1a}
\end{eqnarray}
where the first term is the signature of accretion and the second term attributes to outflow. As the outflow starts from just above $z=0$ surface, within the inflow region itself, we make a reasonable hypothesis that within the prescribed disk region the variation of the dynamical flow parameters with $z$ is much less than that with $r$, allowing us to choose ${\partial A}/{\partial z} \approx O (A/z)$; for any parameter $A$. Strictly speaking, the weak variation with $z$ ensures that the outflow originating from the mid-plane of the disk does not disrupt the disk structure, and thus allowing a smooth accretion flow towards the black hole. Thus Eqn. (\ref{1a}) then reduces to 
\begin{eqnarray}
\frac{1}{r} \frac{\partial}{\partial r} (r \rho v_r) \,+ \, \frac{\rho v_z}{z}=  0.
\label{1b}
\end{eqnarray}

(b) Radial momentum balance:
\begin{eqnarray}
v_r \frac{\partial v_r}{\partial r} \,+ \, v_z \frac{\partial v_r}{\partial z} \, - \, \frac{\lambda^{2}}{r^3} \, + \, F_{Gr} \, + \, \frac{1}{\rho} \frac{\partial P}{\partial r} \,  = \, 0,
\label{2a}
\end{eqnarray}
where $F_{Gr}$ is radial component of the gravitational force. As discussed earlier, here $F_{Gr}$ is the radial force 
corresponding to the PNP in cylindrical coordinate system given by Ghosh \& Mukhopadhyay (2007) containing the information of spin of the black hole. With 
the similar argument as above the equation then reduces to 
\begin{eqnarray}
v_r \frac{\partial v_r}{\partial r} \,+ \, v_z \frac{v_r-v_{r0}}{z} \, - \, \frac{\lambda^{2}}{r^3} \, + \, F_{Gr} \, + \, \frac{1}{\rho} \frac{\partial P}{\partial r} \,  = \, 0,
\label{2b}
\end{eqnarray}
where $v_{r0}$ is the radial velocity at the mid-plane of the disk.

(c) Azimuthal momentum balance:
\begin{eqnarray}
\frac{1}{r^2} \frac{\partial}{\partial r} \left(
r^2 \rho v_r v_\phi \right)
+ \frac {\partial}{\partial z} \left(\rho v_\phi  v_z \right) = 0,
\label{3a}
\end{eqnarray}
where $v_\phi$ is the azimuthal velocity of the flow. The first term of this equation signifies the 
radial transport of the angular momentum, while the second term describes the extraction of 
angular momentum due to mass loss in the outflow. If we consider that the net angular momentum extracted 
by the outflow be $\lambda_j$, 
and the remaining angular momentum retained by the disk $\lambda_d$, then total angular momentum 
$\lambda=\lambda_j +\lambda_d$ can be assumed to remain constant 
throughout the flow within our predefined disk-outflow region by the virtue of an inviscid flow. 
Therefore,  
\begin{eqnarray}
\lambda= {\rm constant}.
\label{3b}
\end{eqnarray}

(d) Vertical momentum balance:
\begin{eqnarray}
v_r \frac{\partial v_z}{\partial r} \, + \, v_z \frac{\partial v_z}{\partial z} \, +  \, F_{Gz} \, + \, \frac{1}{\rho} \frac{\partial P}{\partial z} \, = \, 0,
\label{4a}
\end{eqnarray}
where $F_{Gz}$ is the vertical component of the gravitational force, described by Ghosh \& Mukhopadhyay (2007). Following previous 
arguments this reduces to
\begin{eqnarray}
v_r \frac{\partial v_z}{\partial r} \, + \, \frac{v^{2}_z}{z} \, +  \, F_{Gz} \, + \, \frac{1}{\rho} \frac{P-P_0}{z} \, = \, 0,
\label{4b}
\end{eqnarray}
where $P_0$ is the pressure of the flow at the mid-plane of the disk.  
If there is no outflow, then  
$v_z = 0$, and Eqn. (\ref{4b}) reduces to the well 
known hydrostatic equilibrium 
condition in the disk, and the hydrostatic disk scale-height can be obtained. Similarly, one can customarily extend the 
vertical momentum equation to compute the disk scale-height when there is an outflow coupled with the disk. Thus we will use Eqn. (\ref{4b}) 
to obtain the scale-height of the disk-outflow coupled system. We can reasonably assume that at height $h$ (i.e. at $z$=$h$), the pressure 
of the disk is much less 
compared to that at the equatorial plane to prevent any disruption of the disk, permitting a steady structure of accretion flow. The 
variation of density 
along $z$ direction 
is approximately kept constant ($\rho_0 \, \sim \, \rho$) which in turn means that the disk-outflow coupled region is weakly stratified. Considering 
the above facts, 
Eqn. (\ref{4b}) is simplified to obtain the scale-height as 
\begin{eqnarray}
v_r|_h \frac{\partial v_z}{\partial r}\bigg |_h \, + \, \frac{v^{2}_z |_h}{h} \, +  \, F_{Gz}|_h \, - \, \frac{P_0}{h \rho} \, = \, 0,
\label{4c}
\end{eqnarray}
where $h$ is thus the the root of the above transcendental equation. 
Thus 
Eqns. (\ref{1b}), (\ref{2b}), (\ref{3b}) \& (\ref{4c}) will simultaneously have to be solved to obtain the 
dynamics and the energetics of the accretion-induced outflow. 

\section{Solution procedure and the disk-outflow surface}

We assume that the accretion flow follows an adiabatic equation of state $P=K \rho^{1+1/n}$ 
where $n=1/(\gamma-1)$, $n$ and $\gamma$ are the 
polytropic and 
the adiabatic indices respectively. Note that constant $K$ carries the information of entropy (e.g. Mukhopadhyay 2003) of the 
flow. Thus for an adiabatic flow, $c_s =(\gamma \frac{P}{\rho})^{1/2}$. 
In an earlier work (GM09), a reasonable assumption was made that $h \sim r/2$, which can be approximately accepted 
for a geometrically thick, 2.5-dimensional disk structure. Presuming that the outflow velocity is not likely 
to exceed the sound speed at the disk-outflow surface (the outer boundary of the accretion flow in the $z$ direction), we propose 
a simplified relation between $c_s$ and $v_z$ as 
\begin{eqnarray}
v_z \, \lsim \, 2 (\frac{z}{r}) \, c_s.
\label{5a}
\end{eqnarray}

Hence, at $z=h \, \sim r/2$, we obtain  
\begin{eqnarray}
v_z \, \lsim  \, c_s.
\label{5b}
\end{eqnarray}
This implies that the outflow can ideally come out off the disk surface which can appropriately be termed as the sonic surface 
in the vertical direction. With this notion in mind let us generalize this particular scaling between $v_z$ and $c_s$, pertaining to our 
formalism as 
\begin{eqnarray}
v_z \, = \, {\imath} \, (\frac{z}{r})^{\mu} \, c_s, 
\label{5c}
\end{eqnarray}
where $\imath$ and $\mu$ are the constants which will be determined by substituting Eqn. (\ref{5c})  
in our model conservation equations described in \S 2. The index $\mu$ measures the degree of subsonic nature of the 
vertical flow within the disk-outflow coupled region (i.e. in between the mid-plane  and the surface of the accretion flow). 

Generalizing the procedure adopted in  earlier works for 1.5-dimensional disks (Chakrabarti 1996; Mukhopadhyay 2003; Mukhopadhyay \& Ghosh 2003), 
we solve 
Eqns. (\ref{1b}), (\ref{2b}), (\ref{3b}) \& (\ref{4c}). Using Eqn. (\ref{5c}) along with Eqn. (\ref{3b})
and combining  Eqns. (\ref{1b}) \& (\ref{2b}), we obtain 
\begin{eqnarray}
\nonumber
\frac{\partial v_r}{\partial r}= \frac{\frac{\lambda^2}{r^3} - F_{Gr} + \frac{c^2_s}{r} + 
\frac{\imath}{z} (\frac{z}{r})^{\mu} c_s (v_{r0} - v_r + \frac{c^2_s}{v_r})}{v_r-\frac{c^2_s}{v_r}} = \frac{N}{D}.\\
\label{5d}
\end{eqnarray}
Equation (\ref{5d}) shows that to guarantee a smooth solution at the 
``critical point", $N = D = 0$. From the critical point 
condition we obtain $v_{rc} = c_{sc}$, at $r=r_c$. 
Here subscript $c$ is referred to critical point. The radius $r_c$ is also called the ``sonic radius'' since 
no disturbance created within this radius can cross the radius (also known as sound 
horizon) and escape to infinity. 
Conditions at critical/sonic radius give 
\begin{eqnarray}
\nonumber
v_{rc}  =   c_{sc} =  -\frac{\imath}{2 z} (\frac{z}{r_c})^{\mu} c_{s0c} r_c + \biggl[\biggl\{\frac{\imath}{2 z} \biggl(\frac{z}{r_c}\biggl)^{\mu} c_{s0c} r_c\biggl\}^2 \\+ r_{c} F_{Grc} - \frac{\lambda^2_c}{r^2_c}\biggr]^{1/2}, 
\label{5e}
\end{eqnarray}
where $c_{s0c} = v_{r0c} = (r_{c} F_{Gr0c} - {\lambda^2_c}/{r^2_c})^{1/2}$, is the sound speed or the radial velocity of the flow at the 
sonic radius in the disk mid-plane ($z=0$). $F_{Grc}$ is the radial gravitational force at the sonic radius and $F_{Gr0c}$ is the corresponding 
value in the disk mid-plane.

Now at sonic location $\frac{\partial v_r}{\partial r}=0/0$. Hence by applying  l'Hospital's rule, Eqn. (\ref{5d}) reduces to
\begin{eqnarray}
\nonumber
\frac{\partial v_r}{\partial r}\bigg |_c =-\biggl[2 n \frac{v_{rc}}{c_{sc}} \biggl(\frac{-{\cal B}+\sqrt{{\cal B}^2-4\cal{A}\cal{C}}}{2\cal{A}}\biggr) + \frac{v_{rc}}{r_c}\\+\frac{\imath}{z} \biggl(\frac{z}{r_c}\biggr)^{\mu} c_{sc} \biggr],
\label{5f}
\end{eqnarray}
where ${\cal A} = {\cal F}_1 (r, z)|_c$, ${\cal B} =  {\cal F}_2 (r, z)|_c$ and ${\cal C} = {\cal F}_3 (r, z)|_c$, are complicated functions of $r$ and $z$ at sonic location; 
${\cal F}_1$, ${\cal F}_2$ \& ${\cal F}_3$ have been explicitly given in the appendix. Equations (\ref{5d}) and (\ref{5f}) are 
then solved with an appropriate boundary condition to obtain 
$v_r$ and $c_s$ as functions of $r$. The value of specific angular momentum in our flow always remains constant and 
is same as $\lambda_c$, the value at sonic radius, by virtue of Eqn. (\ref{3b}).

Until now, we have emphasized on velocity profiles of the accretion-induced outflow at any arbitrary $z$. We have, however, 
mentioned before that the intrinsic coupling of inflow and outflow is confined within a specified region, from where 
the outflow emanates. The disk-outflow inter-correlated region is bounded vertically from the mid-plane ($z=0$) to an upper 
surface above which any inflow of matter ceased to exist. We aim at precisely investigating the nature and dynamics of the flow at different 
layers within this disk-outflow
coupled region. Therefore, we first need to calculate the scale-height of the accretion flow, based on our proposed model, and obtain the 
disk-outflow surface (upper boundary). It is to be noted that the transcendental  Eqn. (\ref{4c}) cannot be independently used to calculate the scale-height $h$ 
or its general variance with $r$,  
for unknown variables $v_r$ and $c_s$. Nevertheless, we can easily obtain the scale-height $h$ at $r_c$, say $h_c$, from Eqn. (\ref{4c}) 
as radial velocity and sound speed at the sonic location are known. Thus at $r_c$, Eqn. (\ref{4c}) simplifies to 
\begin{eqnarray}
\nonumber 
{\imath} \biggl(\frac{h_c}{r_c}\biggr)^{\mu} c_{sc} |_h \biggl[{\imath} \biggl(\frac{h_c}{r_c}\biggr)^{\mu} && \frac{c_{sc} |_h}{h_c} - \mu \frac{c_{sc} |_h}{r_c}  + \left(\frac{\partial c_s}{\partial r}\right)_c\bigg |_{h} \biggr]\\ 
&&+ F_{Gzc}|_h -\frac{c_{s0c}^2}{\gamma h_c} = 0, 
\label{6a}
\end{eqnarray}
where ${\partial c_s}/{\partial r}|_{ch}$ is ${\partial c_s}/{\partial r}$ at the sonic location in the plane with scale-height $h$. Then $h_c$ obtained from Eqn. (\ref{6a}) can be 
inserted into Eqn. (\ref{5f}) to obtain ${\partial v_r}/{\partial r}|_c$ at scale-height $h$. 

The disk scale-height $h$ without any outflow is seen to be linearly increasing with $r$ (approximately by 
dimensional analysis). For a 2.5-dimensional disk-outflow system, it seems that 
$h$ and $r$ can also be linearly connected, based on the order of magnitude analysis (GM09). However,
unlike the thin disk, strong gas pressure gradient in the disk-outflow coupled system 
leads to a geometrically thick 2.5-dimensional disk. Therefore, for the present purpose we make 
a generalized scaling of $h$ with $r$ as 
\begin{eqnarray}
h \, \sim \, \delta \, r,
\label{6b}
\end{eqnarray}
where $\delta$ be a dimensionless arbitrary constant or any numerical variable (discussed in detail later). The impressionable choice about the factor 
$\delta$ would be that it should contain precisely the information of the nature of the flow. The only 
physical information we can extract related to the scale-height from our model equations is $h_c$. Thus we rationally demand an approximate 
expression for dimensionless parameter $\delta$ as 
 \begin{eqnarray}
\delta \, \sim \, \frac{h_c}{r_c}. 
\label{6c}
\end{eqnarray}
It is found that the value of $\delta$ circumvents around $1/2$, for the entire range of spin parameter of the black hole, which here acts as a normalization constant.  
Thus eventually, Eqns. (\ref{5d}), (\ref{5f}), (\ref{6b}) \& (\ref{6c}) are combined together to obtain $v_r$ and $c_s$ along the scale-height $h$ for 
an accretion-induced outflow. It also appears that under all these circumstances, and for a physically acceptable flow, the 
constant ${\imath}$ and the index $\mu$ of Eqn. (\ref{5c}) connecting $v_z$ and $c_s$ yield to be $\sim 1$ and $\sim 3/2$ respectively. 

\subsection{Construction of disk-outflow surface}

After establishing the scale-height $h$, we in principle can obtain the profiles of the dynamical parameters at 
different layers of the disk, i.e. at $z={\ell}h$, where $0 \leq {\ell} \leq 1$, within our 
prescribed disk-outflow region. We find that the radial velocity profile $v_r$ along 
all layers of disk for any arbitrary spin parameter exhibits some unusual, yet very interesting behavior. For 
any specific $z$ and spin parameter $a$, $v_r$ attains a negative value at a radius greater than a certain distance $r = r_b$. 
The magnitude of $r_b$ decreases with the increase of $\ell$, which indicates that the positive trait of $v_r$ is greater for lower 
$z$. The negative value of $v_r$ does not necessarily reflect any unphysical behavior. 
If the outflow originates from any particular layer ${\ell} h$ at a radius $r\ge$ $r_b$, it then appears that the coupling 
between the disk and the outflow ceased to exist. It infers that along the layer ${\ell} h$, both accretion and outflow simultaneously occur up to the 
radial distance $r_b$, and beyond which the solution (with negative $v_r$) reflects the truncation of the disk (along the specified layer). 
Let us consider different layers of $z$. As we ascend to layers from a lower to a higher $z$, the truncation of 
the disk-outflow region along these layers occurs at a radius smaller than that of the lower $z$. As accretion is the source of the outflow, we can 
ostensibly conclude that the region from where the outflowing matter originates in the disk shrinks as we go to higher latitudes. Note that 
$r_b$ corresponds to zero $v_r$. 
Therefore, we restrict our study up to the boundary $r_b$. With this insight, we 
simply compute $r_b$ along various layers of the disk and construct a surface in the $r-z$ plane connecting all $r_b$ for different layers. The enclosed region bounded by this 
surface is defined by the positivity of $v_r$, where both the accretion and outflow simultaneously persist and are intrinsically coupled to each other. We attribute the 
outer surface of the region as disk-outflow surface ($h_{surf}$) and above this layer no inflow takes place. The arrows in the diagram shown in Fig. 1 
reveal the direction of flow. The arrow exactly at $r_b$ points vertically upwards, which indicates that just at $r_b$ the flow pattern 
exclusively corresponds to vertical motion and any accretion flow ceased to exist. However, more realistic disk-outflow surface could be visualized with a thick-solid 
line shown in Fig. 1.

%\clearpage

\begin{figure}
\centering
\includegraphics[width=1.0\columnwidth]{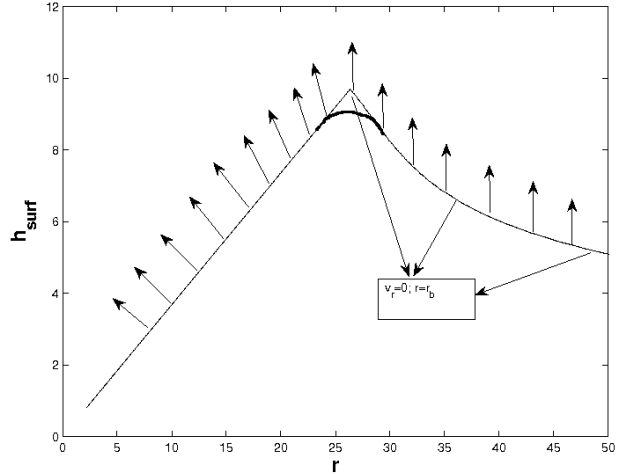}
\caption{
Nature and geometry of the disk-outflow coupled region and the outflow surface. Discussed in detail in \S 3.1.
}
\label{Fig1}
\end{figure}

As in our system we have neglected the viscosity and any radiative loss, and thus the flow is considered 
to be strongly advective, presumably it is 
gas pressure dominated. Notwithstanding, various scattering processes will indeed produce radiation 
in the system, whose contribution 
could be insignificant. Hence, $\gamma \sim 3/2$ is reasonably an appropriate choice in our model 
\footnote{The relationship between $\gamma$ and $\beta$, the ratio 
of gas pressure to the total pressure, is given by $\beta=\frac{6\gamma-8}{3(\gamma-1)}$ (GM09).}. 
Note that this is in essence similar to the choice of the earlier authors who
modeled gas dominated low mass advection dominated accretion flows (e.g. Narayan \& Yi 1994). 
We would also like to clarify that $\gamma=5/3$ corresponds only to the 
case of zero angular momentum. 
Figure 2 shows the profiles of disk-outflow surface for various spin parameters $a$ of the 
black hole. All the input parameters corresponding to each value of $a$ 
are listed in Table 1.
It is seen that the disk-outflow region and the peak of the surface namely, $R_{js}$, shifts to the vicinity of the black hole for a higher spin.

%%%%%%%%%%%%%%%%%%%%%%%%%%%%%%%%%%%%%%%%%%%%%%%%%%%%%%%%%%%%%%%%%%%%%
%\clearpage
%vskip0.2cm
%{\centerline{\large Table-4}}
%{\centerline{\large Values of $r_b$}}
%\begin{center}
%
%vbox{
\begin{table*}[htbp]
%\scriptsize
%{\centerline{\large Table-4}}
%{\centerline{\large Values of $r_b$}}
%\begin{center}
\Large
\centerline{\large Table 1}
\centerline{\large Input parameters for various $a$ }
\begin{center}
\begin{tabular}{ccc}
\hline
\hline
 $a$  & $r_c$  & $\lambda$  \\
\hline
\hline
0.0  & 6.15 & 3.3  \\
0.3 & 5.7 &  2.9  \\
0.5 & 5.5 & 2.6  \\
0.7 & 5.3 & 2.3  \\
0.9 & 5.0 & 1.9  \\
0.95 & 4.4 & 1.6  \\
0.97 & 4.3 & 1.5 \\
0.998 & 4.2 & 1.3  \\
\hline
\hline

\end{tabular}

\end{center}
\end{table*}

%\clearpage

\begin{figure}
\centering
\includegraphics[width=1.0\columnwidth]{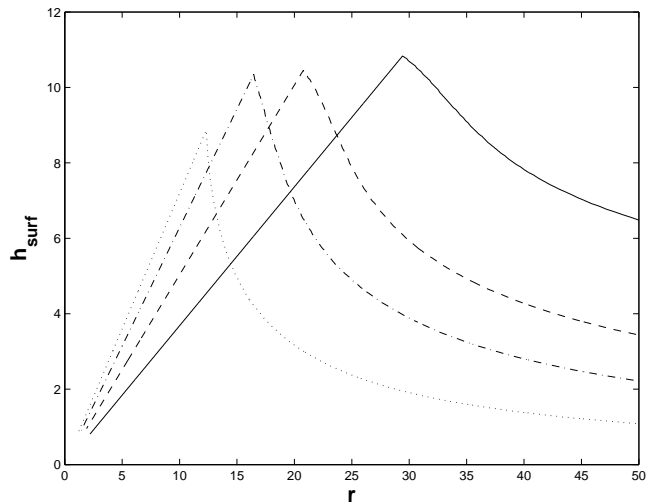}
\caption{
Variation of disk-outflow surface with radial coordinate for different spin of the
black hole. Solid, dashed, dot-dashed and dotted curves are for $a=0, 0.5, 0.9, 0.998$
respectively. $\gamma = 1.5$.
 }
\label{Fig2}
\end{figure}

The marginally stable orbit in a Keplerian accretion flow defines the inner edge of the disk (with zero torque 
condition). However the orbit is more or less an artifact of the flow. 
For a transonic advective disk (as is the present case), an apriori definition of inner edge is somewhat fuzzy, due to which we can pretend to choose that the disk extends 
to the black hole horizon. In presence of an outflow intrinsically coupled to the disk, there is supposed to be an inner boundary beyond which any 
outflow and then jet will ceased to exist. We define this inner boundary of the disk-outflow coupled region in explaining the energetics of the flow. 
Also, with the increase of the spin of the black hole, it is seen that the disk-outflow coupled region shrinks 
considerably, attaining a steeper nature. Literally speaking, the 
spin of the black hole directly influences the nature and region of the outflow. The tendency of the outflow region to get contracted to 
the inner radius suggests that the outflow is more likely to exuberantly emanate from inner region of the disk for rapidly 
spinning black holes which have more gravitating power. Therefore, with the increase in spin, the disk-outflow region becomes more dense and more susceptible to eject the matter with a greater power. 

\section{Dynamics and nature of the flow}

As the system is gas pressure dominated and strongly advective, it is more receptive to strong outflows. Figures 3 and 4 describe the variation of flow parameters as functions of radial coordinate $r$ along the disk-outflow surface for various spin of the black hole. With the increase of the spin of the black hole, $v_r$ 
increases at the inner region of the disk. It is seen that along the disk-outflow surface, $v_r$ becomes zero beyond a certain distance $R_{js}$, which is also termed as zero $v_r$ surface illustrated in \S 3.1, which is explained it detail in the next section in order to understand the outflow power. The 
zero $v_r$ surface extends upto more inner region of the disk with the increase of spin of the black hole, indicating the fact that outflows and then jets originate from more inner region of the 
disk for rapidly spinning black holes. 
The increase of $c_s$ with spin (Fig. 4) indicates that the temperature of the disk-induced outflow is higher for rapidly spinning black holes. It is found from Fig. 4 that the maximum temperature of the accretion-induced outflow varies from $\sim 10^{11}$ to $10^{12} K$ corresponding to zero to maximal
spin of the black hole. The truncation of the curves in the inner region symbolizes the inner boundary of the disk-outflow region as mentioned in \S 3.1.

%\clearpage

\begin{figure}
\centering
\includegraphics[width=1.0\columnwidth]{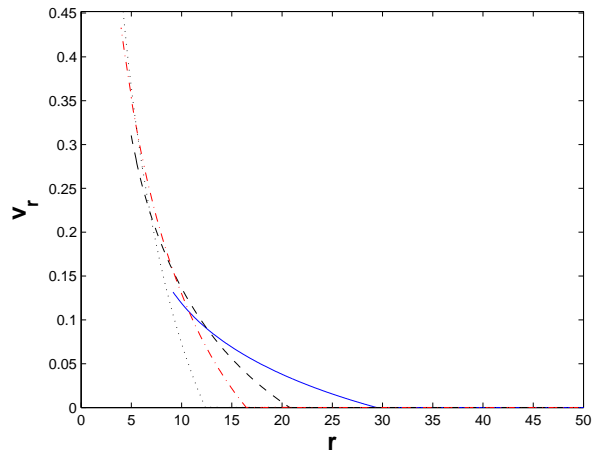}
\caption{
Variation of radial velocity with radial coordinate.  Solid, dashed, dot-dashed and dotted curves are for $a=0, 0.5, 0.9, 0.998$ 
respectively. Other parameter $\gamma = 1.5$.
 }
\label{Fig3}
\end{figure}

%\clearpage

\begin{figure}
\centering
\includegraphics[width=1.0\columnwidth]{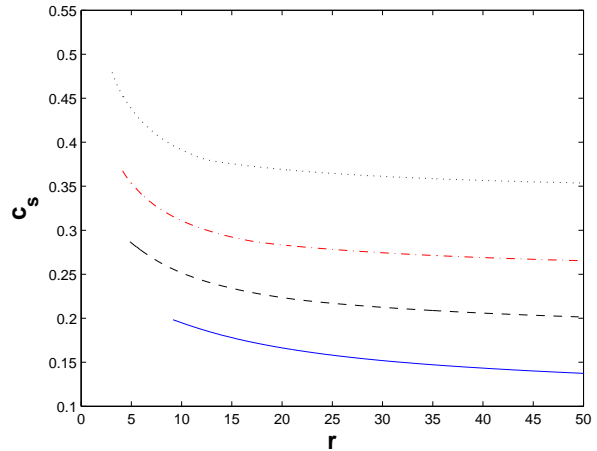}
\caption{
Same as that of Fig. 3 but variation of sound speed.  
 }
\label{Fig4}
\end{figure}

\section{Energetics of the flow}

Accretion by  a black hole is the primary source of energy of the mass outflow from the 
inner disk region. The energetics of the accretion-induced outflow are mainly attributed to the 
mass outflow rate and the power extracted by the outflow from the disk. The derivation 
of the mass outflow rate had been elaborated in G10. We follow the same procedure here, however 
including the spin information of the black hole. The mass outflow rate is given by 
\begin{eqnarray}
\dot{M}_j (r) \, = \, - \int 4 \pi r \rho (h_{surf}) v_z (h_{surf}) \, dr + c_j,
\label{5a}
\end{eqnarray}
where the constant $c_{j}$ is determined by an appropriate boundary condition. 
From the adiabatic conditions, $\rho$ can be written in terms of $c_s$ which 
is already determined as a function of $r$. The corresponding proportionality constant is 
determined at a radius, outside which the contribution to the mass outflow rate 
is negligible ($v_z=0$). However, the total mass accretion rate ${\dot M}$
(which is sum of the inflow rate and the outflow rate)
can be obtained by integrating the continuity equation along the radial 
and vertical directions.  Hence, the constant is computed easily by supplying 
the values of $v_r$ and $c_s$ at that radius in the expression of ${\dot M}$.
As we have discarded any possible small amount of the outflow at that
radius, the actual value of the constant, and hence the outflow power, could 
be slightly over estimated.

$\dot{M}_j (r)$ in Eqn. (\ref{5a}) refers to the 
rate at which the vertical mass flux ejects from the disk-outflow surface. 
Figure 5 shows the variation of $\dot{M}_j$ profiles with the spin of the black hole. It is seen that with the increase of spin 
of the black hole, $\dot{M}_j$ increases. All the profiles have been shown considering a supermassive black hole 
of mass $\sim 10^{8} M_{\odot}$ with a mass accretion rate at infinity, $\dot M \sim 10^{-2} \dot M_{Edd}$, where $\dot M_{Edd}$ is the Eddington mass accretion rate 
$\sim 1.44 \times 10^{25} gm/s$. Low mass accretion rate (sub Eddington accretion flow) is in conformity with our gas pressure 
dominated advective disk paradigm.
It is found that $\dot{M}_j$ increases with the increase of mass of the black hole and mass accretion rate as well. The truncation of the curves at an inner radius indicates the inner boundary of the disk-outflow coupled region, explained in detail in the next paragraph.  

%\clearpage

\begin{figure}
\centering
\includegraphics[width=1.0\columnwidth]{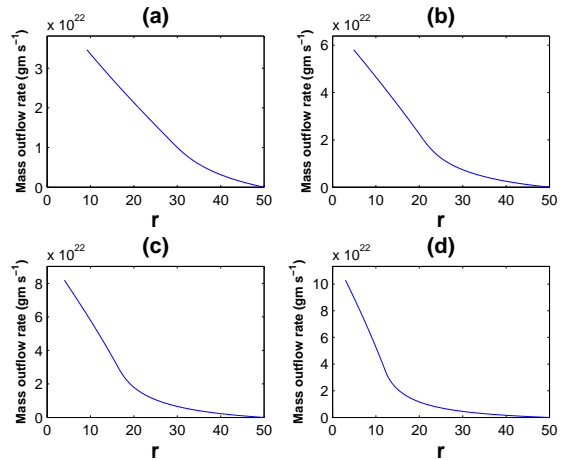}
\caption{
Variation of mass outflow rate as a function of radial coordinate, when 
(a) $a$ = 0, (b) $a$ = 0.5, (c) $a$ = 0.9, (d) $a$ = 0.998. Other parameter $\gamma = 1.5$.
 }
\label{Fig5}
\end{figure}

In computing the power extracted by the outflow from the disk, we follow the same procedure as in G10. Thus the power of the outflow is given by 
\begin{eqnarray}
\nonumber
P_j (r) = \int 4 \pi r \biggl[\biggl(\frac{v^2}{2} + \frac{\gamma}{\gamma-1} 
\frac{P}{\rho} + \phi_G  \biggr) \rho v_z \biggr] \bigg |_{h_{surf}} \, dr, \\
\label{5b}
\end{eqnarray}
which is the total power removed from the disk by the outflow along the disk-outflow surface. In measuring the net power of any 
astrophysical jet, it appears that the computed power $P_j$ will then be the initial power of the jet. Figure 6 depicts the 
variation of the power $P_j$ with $r$ for various spin parameters of the black hole. If we meticulously investigate the 
nature of the power profiles, we observe that when the radial distance is less than a certain value, $P_j$ begins to decrease. The decreasing trend of $P_j$ infers that the characteristic flow is bounded (integrand of Eqn. (\ref{5b}) carrying the information of the vertical energy flux becomes negative). 
%\clearpage

\begin{figure}
\centering
\includegraphics[width=1.0\columnwidth]{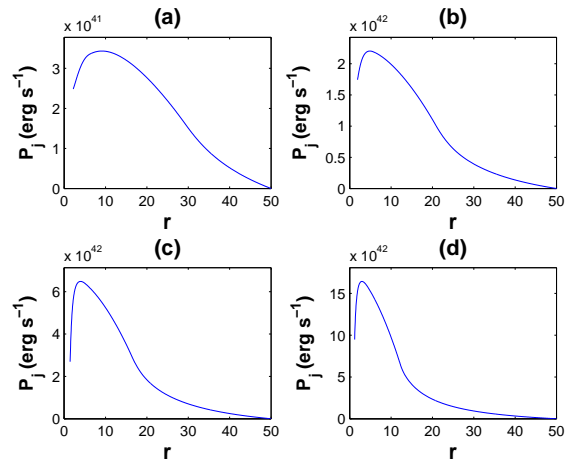}
\caption{
Variation of outflow power as a function of radial coordinate, when (a) $a$ = 0, (b) $a$ = 0.5, (c) $a$ = 0.9, (d) $a$ = 0.998. Other parameter $\gamma = 1.5$.
 }
\label{Fig6}
\end{figure}

This occurs owing to the fact that in the extreme inner region of disk, due to its strong gravitating power the starved black hole sucks 
all of the matter in its sphere of influence, even if there is any outflow emanating from the disk. We identify this inner transition 
radius as $R_{jt}$, beyond which no outflow occurs. We attribute $R_{jt}$ as the inner boundary of the disk-outflow region. 
In all of the previous profiles, the said truncation of the curves is ascribed to $R_{jt}$. We describe the dynamical variables in our study within the region between an 
outer boundary and $R_{jt}$, and within this prescribed region strong outflows are most plausible to originate, intrinsically coupled to the disk. 
The power profile shows that with the increase in $a$, $P_j$ increases as well as $R_{jt}$ gradually shifts to the vicinity of the black hole,
 indicating the fact that outflow 
region moves further inward. We show the variation of total $P_j$ extracted from the disk-outflow region with the spin of the black hole in Fig. 7a. 
Considering a black hole of mass $\sim 10^{8} M_{\odot}$, as 
seen in AGNs and quasars, accreting with $\dot M \sim 10^{-2} \dot M_{Edd}$, it is seen that 
$P_j \sim 10^{41}$ $erg/s$ for $a=0$. For a maximally spinning black hole ($a=0.998$) with same parameters, 
the computed  $P_j \sim 10^{43}$ $erg/s$. Thus, 
there is an increase of two orders of magnitude of $P_j$ with the increase of spin of 
the black hole from $0$ to $0.998$. In an 
earlier work, Donea \& Biermann (1996) showed that the power extracted by the 
outflow/jet from the disk increases with $a$.
However, they did not compute the power extracted from the disk explicitly. The numerical
simulations by De Villiers et al. (2005), in a different accretion paradigm including magnetic field, 
also concluded that the jet efficiency is possible to
increase with the increase of spin of the black hole from $0$ to $0.998$. 
Figure 7b shows the 
variation of $R_{jt}$ with $a$. In describing the disk-outflow surface in \S 3.1, we have articulated the impact of spin on the disk-outflow coupled region. The geometry of 
the surface for a particular $a$ is identified with a parameter $R_{js}$ (see Fig. 2). 
We show exclusively the variation of $R_{js}$ with the spin of the 
black hole in Fig. 7c. which too reveals that the fast rotating black hole retracts the outflow region towards it; a sign of a pure relativistic 
gravitation. 

%\clearpage

\begin{figure}
\centering
\includegraphics[width=1.0\columnwidth]{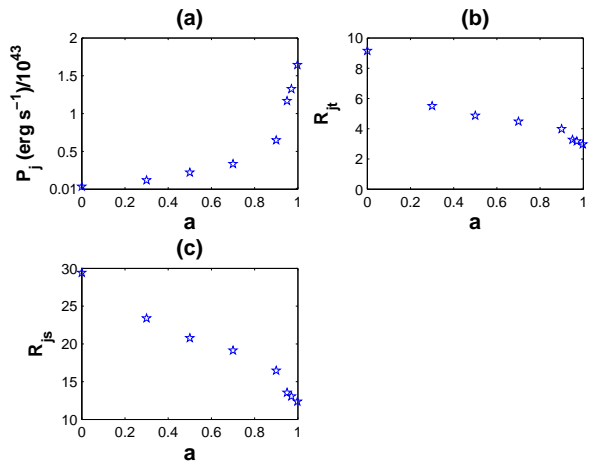}
\caption{
Variations of (a) net outflow power in the units of $10^{43}$ erg s$^{-1}$, (b) inner transition radius of the outflow, (c) peak of 
disk-outflow surface, as functions of spin of the black hole.
 }
\label{Fig7}
\end{figure}

\section{Discussion}

Blandford-Znajek process  (Blandford \& Znajek 1977) is still one of the most promising mechanisms to drive powerful 
jets in AGNs and XRBs. Although the 
exact mechanism of formation of the jet in the vicinity of the black holes is still elusive, and whatever might be the reason for the origin of jet, the said work is 
significant mostly due to two underlying reasons: (1) extreme gravity is indispensable to effuse strong unbounded flows in the 
vertical direction from the inner region of the accretion disk, (2) the spin of the 
black holes, which is purely a relativistic effect, is essential to power strong outflows and jets.  

In the present study, we have neither laid importance to the nature of the outflow nor invoked any aspect for the origin of outflows or jets. Indeed, the distinctive or definitive understanding 
of the formation of strong outflows or jets is till unknown. Notwithstanding, the most distinguishable and obvious picture in this case is that outflows and jets 
observed in AGNs and XRBs can only originate in an accretion powered system, where the accretion of matter around relativistic gravitating objects like black holes acts as a source, and outflow and then jet takes the form of one of the possible sinks (the other sink is the central nucleus). The dynamics of the outflowing matter should then be intrinsically coupled to the accretion dynamics macroscopically through the fundamental laws of conservation (of matter, momentum and energy). The outflow is unbounded and 
the total energy just at the base of the outflow should be positive. The accretion flow should be bounded in the vicinity of the central 
object as the central potential is attractive. However, the strength of the unbounded flows in the form of jets is distinctly proportional to the attractiveness of the central gravitational potential field. 
This paradox is well manifested in the observed universe, as relativistic jets
are more populated around extreme gravitating objects like black holes. 
Also noticeably, length scale of jets increases from microquasars to quasars, which are supposedly 
harboring stellar mass and supermassive black holes respectively. 

Thus in any theoretical modeling of the accretion and outflow, it can be presumably argued that the mathematical 
equations governing the dynamics of the inflow and outflow should inherently be correlated and evolved self-consistently without any 
ad hoc proposition, as accretion and outflow should not be treated as dissimilar objects. Second, the relativistic gravitational effect of the black hole should be incorporated, as the nature of gravity is the cornerstone to both the accretion and the unbounded outflow. To 
capture this essential physics, in our present study to understand the connection between disk and outflow, we have incorporated the general relativistic effect of the 
spinning black hole through a pseudo-Newtonian approach. Although pseudo-Newtonian formalism is an approximate method to mimic the space-time geometry of the 
Kerr black hole, yet it captures the important salient features of the corresponding metric, and thus can be used to examine the nature of outflows from the inner region of the disk. 

In \S 2 we have described the general disk-outflow coupled 
hydrodynamic equations in the 
 inviscid limit  following GM09. The necessity to simplify our model equations for an inviscid flow is discussed in \S 2. 
Although GM09 explored the 2.5-dimensional accretion-induced outflow for a fully viscous system, they used a self-similar approach in order to solve the necessary partial coupled differential equations. In addition, they neglected the most indispensable effect of relativistic gravitation or,  precisely, the effect of spin of the black hole. In the present paper, we have solved the disk-outflow model equations in a more general 2.5-dimensional
paradigm, while trying to limit our assumptions to the least possible extent theoretically. One of the important premises we have made is the relationship between $v_z$ and $c_s$,  which we have established empirically. We have not used the height integrated equations which are mostly valid in the circumstances where the dynamical fluid parameters are 
likely to be independent of $z$. Without presuming the fact that the outflow originates from the surface of the disk, as most of the authors do, we have logically constructed a disk-outflow surface with properly defined boundary conditions. One of the most important computations we have done is to evaluate the mass outflow rate and the power of the outflow extracted from the disk self-consistently in the inviscid limit, unlike the previous works (e.g. Donea \& Biermann 1996; Blandford \& Begelman 1999; Xie \& Yuan 2008). We have found that the spin of the black hole 
plays a crucial role in determining the structure, dynamics and the energetics of the outflow coupled to the disk. 
With the increase of the spin, the outflow region extends further inward and then the 
disk-outflow region shrinks and compresses (see Fig. 2). As a result, the outflow and then
any plausible jet is likely to eject out from an extreme 
inner region of the flow around a rapidly spinning black hole with a greater efficiency. Note
that the higher spin results in the system to get more compressed with a greater outflow power.
Therefore, the efficiency of outflow and jet is 
directly related to the disk scale-height and hence the disk-outflow surface. 
Previously, in a different
context, Mckinney \& Gammie (2004), while examining the electromagnetic luminosity of a
Kerr black hole, assumed the ratio of the disk height to the radius ($h/r$) constant,
irrespective of the black hole spin. However, as seen in the present work, 
the constant $h/r$ for different spin of the  black hole
may not be an obvious choice.

The power extracted by the outflow from the disk not only 
depends directly on the mass of the black hole and the initial mass accretion rate of the flow, but also on the spin of the black hole. With our model, keeping the  black hole mass 
and the accretion rate the same, the power of the outflow increases with the spin of the black hole and the computed power differs in two orders of magnitude between non-rotating and maximally rotating black holes. We have restricted our study vertically up to the region where the inflow and the outflow are least coupled, i.e. the disk-outflow surface. Above this surface 
accretion ceased to exist and probably the outflow gets decoupled from the disk, accelerates and eventually forms relativistic jet. The modeling of the astrophysical jet is altogether a different issue and is beyond the scope of the present work. Nevertheless, it can be effectively argued that in modeling the dynamics of the jet, the 
computed outflow power may serve as an initial power fed to the jet. Thus, the dynamics and the energetics of the jet will eventually be related to the black hole spin. 

According to the unification scenario, it is possible to device a single
astrophysical scenario, which can
broadly explain the observed multitude of different types of AGNs (see e.g.
Antonucci 1993; Urry \& Padovani 1995). Flat Spectrum Radio Quasars (FSRQs) and BL Lacs are probably
the two most active types of AGNs which are collectively referred to blazar. It
is observed that BL Lacs are relatively low luminous than FSRQs. Although both of these
galaxies show high energy emissions, their spectral properties are different
(Bhattacharya et al. 2009) which indicate that they are different
source classes. According to the unification scheme, for FSRQs the line-of-sight is
almost parallel to the jet and, hence, strong relativistic Doppler beaming of the jet
emission produces highly variable and continuum dominated emission. As one moves
away from the jet axis, the central continuum emission falls and the nucleus looks
like a FR-II galaxy. Similarly, BL Lacs are considered to be a sub class of FR-I
galaxies whose line-of-sight is almost parallel to the jet axis. The present work suggests
that the total mechanical power of an outflow proportionately increases with the spin of the central
supermassive black hole; higher the spin, stronger is the outflow. It is
reasonable to consider that strong outflow can lead to a strong jet, and hence, one
can expect to observe higher luminosity. Therefore, the work suggests that BL Lacs are slow rotators than FSRQs.

For the theoretical formulation of the disk-outflow coupling even with approximations, one needs to be very thoughtful for the proper foundation of the model which essentially needs to solve hydrodynamic or magnetohydrodynamic conservation equations in presence of strong gravity. The flow 
parameters vary in 2.5-dimension and are coupled to each other. Limited observational inputs put irremediable constraint on the boundary conditions as well 
as the fundamental scaling parameters, governing the coupled dynamics of the accretion and outflow. The inadequacy of an effective mathematical tool to handle partial 
coupled differential hydrodynamic equations for compressible flow motivates us to invoke approximations and assumptions. Despite of this fact, one needs to explore 
the possibilities to examine the accretion-induced outflow. The question then arises: what are the valid assumptions and to which extent 
they can be considered? In the present work, we have addressed this question in the following way.\\
(1) As the extreme gravity is the most important aspect to effuse jet from the disk, we have incorporated its effect through a pseudo-general-relativistic potential.\\
(2) The disk and outflow should not be treated as dissimilar objects, and hence their correlated-dynamics should be essentially governed by the conservation laws. 
The energetics of the accretion-induced outflow would then be evaluated self-consistently as is shown in our work. \\
(3) Any unbounded flow in the form of outflow is more plausible to emanate from a hot, puffed up region of the accretion flow (may be low/hard state of the black hole). Hence we have formulated our model in a 2.5-dimensional, strongly advective paradigm, surpassing the simplicity of height integration. \\
(4) We have approximated our system to an inviscid limit, whose reason has been argued in \S 2. Nevertheless, in any future work, viscosity should be incorporated into the flow to 
make it more realistic. \\
(5) We do not account for the mechanism of formation of strong jets, like due to magnetic field or strong radiation pressure, as the definitive mechanism of 
the formation of jet is still unknown. Indeed, it is beyond the scope to solve magnetohydrodynamic equations in the present scenario. However, assuming 
that the outflow resides, and is coupled to the disk, the governing conservation equations of matter, momentum and energy should be treated accordingly. \\
(6) Power of the outflow and jet is expected to directly depend on the spin of the black hole. 
The spin, which is the signature of general relativity, can make an impact 
on the nature of the observed AGN classes, and thus the spin of the black has been 
incorporated in our study. As spin of the black hole has a 
direct impact on the flow parameters, we obtain different outflow power for different spin. 

What can be the most defined parameter for the accretion powered
system --- mass accretion rate, mass of the central star/black hole
or the spin of the central star/black hole? Our analysis has shown
that the power extracted by the outflow from the disk
proportionately increases with the spin of the black hole. It
infers that if extreme gravity is essential to power the jet,
then the strength and the length-scale of the observed
astrophysical jets directly depend on the spin of the black hole.
If it is believed that the distant quasars harbor massive black
holes of same mass scale and accrete matter with similar rate, perhaps, spin be the guiding parameter for the different observed AGN
classes. We can end with the specific question: are the
observed AGN classes astrophysical laboratories to measure
the spin of the supermassive black holes? \\

This work is supported by a project, Grant No. SR/S2HEP12/2007, funded by DST, India. 
The authors would like to thank the referee for making 
important comments which helped 
to prepare the final version of the paper. \\ 

\noindent{\bf APPENDIX}

Equation (\ref{5f}) consists of complicated terms, where ${\cal A}, {\cal B}$  and ${\cal C}$ are given by following equations
$$
%\begin{equation}
{\cal A} \, = \, 4 n+2, 
\eqno(A1)
%\end{equation}
$$
$$
%\begin{equation}
%\nonumber
{\cal B} = \biggl(F_{Grc}-\frac{\lambda_{c}^{2}}{r_{c}^3}\biggr) \biggl(\frac{1}{2 n}+1\biggr) \frac{1}{v_{rc}} + \frac{v_{rc}}{r_c} \biggl(1-\frac{1}{2n}\biggr)  
$$
$$
-\frac{\imath}{z} \biggl(\frac{z}{r_c}\biggr)^{\mu} v_{r0c} \biggl(\frac{1}{n}+1\biggr) + 
2 \frac{v_{rc}}{r_c} + 
2 \frac{\imath}{z}\biggl(\frac{z}{r_c}\biggr)^{\mu} v_{rc},
\eqno(A2)
%\end{equation}
$$
and
$$
%\begin{equation}
%\nonumber
{\cal C} = \biggl(F_{Grc}-\frac{\lambda_{c}^{2}}{r_{c}^3}\biggr) \frac{1}{2 nr_c} + \frac{v^{2}_{rc}}{2 n r_c^2} - \frac{\imath}{2 n z} \biggl(\frac{z}{r_c}\biggr)^{\mu} v_{r0c} v_{rc} \biggl[\frac{1}{r_c} + 
$$
$$
\frac{\imath}{z} \biggl(\frac{z}{r_c}\biggr)^{\mu}\biggr] + 
%\nonumber
\frac{1}{2n} \biggl(\frac{\partial}{\partial r} F_{Gr} \bigg |_{c} +3 \frac{\lambda_{c}^{2}}{r_{c}^4}\biggr) + 
\frac{v^{2}_{rc}}{2 n r_c^2} +\mu \frac{\imath}{2 n z} \, \frac{z^{\mu}}{r^{\mu +1}_c}  \, v_{r0c} v_{rc}  \,
$$
$$
 + \frac{\imath}{2 n z} \biggl(\frac{z}{r_c}\biggr)^{\mu} v_{rc} \biggl[\frac{2 n v_{r0c}}{c_{s0c}} \,   \biggl(\frac{-{\cal B}_{0}+\sqrt{{\cal B}_{0}^2-4{\cal A}_{0}{\cal C}_{0}}}{2{\cal A}_{0}}\biggr)  +\frac{v_{r0c}}{r_c}\biggr], 
\eqno(A3)
%\tag{A3}
%\end{equation}
$$
where, \\
$${\cal A}_{0} \, = \, 2 + 4 n,
\eqno(A4)$$ \\
%continuity eqn
$${\cal B}_{0}=\bigl(F_{Gr0c}-\frac{\lambda_{c}^{2}}{r_{c}^3}\bigr) \bigl(\frac{1}{2 n}+1\bigr) \frac{1}{v_{r0c}} + \frac{v_{r0c}}{r_c} \bigl(1-\frac{1}{2n}\bigr) + 2 \frac{v_{r0c}}{r_c}, \eqno(A5)$$ \\
and \\
$${\cal C}_{0}=\bigl(F_{Gr0c}-\frac{\lambda_{c}^{2}}{r_{c}^3}\bigr) \frac{1}{2 n r_c} + \frac{v^{2}_{r0c}}{n r^2_c} + \frac{1}{2n} \bigl(\frac{\partial}{\partial r} F_{Gr0} \big |_{c} +3 \frac{\lambda_{c}^{2}}{r_{c}^4}\bigr).
\eqno(A6)$$

%\include{Appendix}

%%%%%%%%%%%%%%%%%%%%%%%%%%%%%%%%%%%%%%%%%%%%%%%%%%%%%%%%%%%%%%%%%

%%%

%\label{lastpage}

\end{document}